\documentclass[12pt]{iopart}
\usepackage{graphicx}
\usepackage{iopams}
\usepackage{setstack}
\usepackage{bm}
\input{epsf}

\newcommand{\der}[3]{\frac{\partial^{#1}{#3}}{\partial{#2}^{#1}}}

\def\beq{\begin{equation}}
\def\eeq{\end{equation}}
\def\ba{\begin{eqnarray}}
\def\ea{\end{eqnarray}}

\def\ca{\thicksim}
\def\b0{\arrowvert_{\beta=0}}
\def\zz{\zeta^{-1}_{0\{\beta\}}(z)}

\begin{document}
\title{Periodic orbit theory of strongly anomalous transport}

\vskip4mm

\author{Roberto Artuso\footnote[1]{also at Istituto Nazionale Fisica Nucleare, Sezione di Milano, Via Celoria 16, 20133 Milano, Italy} and Giampaolo Cristadoro}
\address{Center for Nonlinear and Complex Systems and
Dipartimento di Scienze Chimiche, Fisiche e Matematiche, Universit\`a
dell'Insubria and Istituto Nazionale di Fisica della Materia,
Unit\`a di Como, Via Valleggio 11, 22100 Como, Italy}
\eads{\mailto{Roberto.Artuso@uninsubria.it}, \mailto{Giampaolo.Cristadoro@uninsubria.it}}
\begin{abstract}
We establish a deterministic technique to investigate transport
moments of arbitrary order. The theory is applied to the analysis of
different kinds of intermittent one-dimensional maps and the Lorentz gas with infinite horizon: the typical appearance of phase transitions in the spectrum of transport exponents is explained.
\end{abstract}

\pacs{05.45.-a }
\maketitle

\section{Introduction}
It is well known that chaotic dynamical systems may exhibit deterministic diffusion: among the simplest examples one may recall chains of 1-d maps \cite{1D}, area preserving maps on the cylinder \cite{2D} and infinite domain billiards, like the Lorentz gas with  finite horizon \cite{LF}. In many cases the mean square displacement (the average being taken with respect to initial conditions) displays a linear asymptotic growth
\beq
\langle (x_t-x_0)^2 \rangle \sim Dt
\eeq
and one may try to devise techniques to compute the diffusion constant $D$. In particular periodic orbits expansions have been employed \cite{POE} to evaluate $D$ for one dimensional maps \cite{PD1}, finite horizon Lorentz gas \cite{PD2} and area preserving sawtooth maps \cite{PD3}. Deterministic transport may also exhibit an anomalous character, with a mean square displacement deviating from a linear behaviour. Typically this is associated to a weakening of chaotic properties, like marginal fixed points \cite{A1}, the presence of regular islands in the chaotic sea \cite{A2}, or the opening of ballistic channels for Lorentz gas models \cite{A3}. Such weakly chaotic regimes are generically quite hard to study, as they miss the fast memory loss that typically characterizes strongly chaotic systems. Besides transport properties this lack of  full chaos deeply modifies other dynamical features: for instance escape processes may be slowed down and time correlation functions acquire power law tails (in view of Green-Kubo formula \cite{GK} this is closely connected to anomalous diffusion). Global properties of the transport process are investigated  through the  asymptotic behaviour of generalized moments
\beq
\langle \mid x_t-x_0\mid^q \rangle \sim t^{\nu(q)}.
\eeq
Gaussian processes yield $\nu(q)=q/2$ for all $q$; recently deviations from a constant slope for $\nu(q)$ have been referred to as ``strong'' anomalous diffusion \cite{V-AD} (the reader may consult this reference for a wide survey of systems falling in this cathegory).
In this paper we will introduce a cycle expansions technique for computing the spectrum $\nu(q) \,\, (q/2 \in \mathbb{N})$ and apply the formalism to a number of examples: different chains of one-dimensional intermittent maps, and infinite domain billiards (Lorentz gas with infinite horizon).

The paper is organized as follows: in section $2$, to make the paper self contained, we review how periodic orbit expansions may be employed to study transport processes, section $3$ briefly explains how full chaos generically results in a gaussian $\nu(q)$, while, finally, section $4$ is devoted to the analysis of weakly chaotic systems yielding a non trivial $\nu(q)$ (parts of the results of this section were anticipated in \cite{rgp}).
\section{Deterministic transport}
Chaotic dynamical systems are characterized by sensitive dependence on initial conditions, so that a meaningful theory must concentrate on statistical properties, rather than on analysis of single trajectories. A central role in such a context is played by the Perron-Frobenius operator $\mathcal{L}$ which acts as:
\begin{equation}\label{Perron}
 [\mathcal{L}\phi](x)=\int_X \rmd y \,\, \delta(f(y)-x) \phi(y)
\end{equation}
(we are considering for simplicity dynamical evolution described by a discrete mapping $f$).
This operator may be interpreted as the evolution on densities (dual picture with respect to Koopman dynamics on observables)
\beq
\rho_{n+1}(x)=\mathcal{L}[\rho_n](x)=\int_X \rmd y \,\, \delta(f(y)-x) \rho_n(y)
\end{equation}
(in particular the invariant density is an eigenvector of $\mathcal{L}$, with corresponding eigenvalue $\lambda=1$).
Notice that the kernel is singular, so that investigation of spectral properties generally requires sophisticated techniques \cite{VB}: we also observe that this operator enjoys the semigroup property:
\begin{equation}
[\mathcal{L}^{m}\phi](x)=\left[ \mathcal{L}^{m-k}[\mathcal{L}^k \phi] \right](x)
\end{equation}
The study of transport properties is conveniently carried out in terms of the generating function:
\begin{equation}\label{genfunc}
G_t(\beta) =\quad \langle \rme^{\beta (x_t-x_0)} \rangle
\end{equation}
when again the average is over a set of initial conditions.
 We also define a  {\em generalized} transfer operator where the singular delta kernel in (\ref{Perron}) is weighted by an appropriate factor (that maintains the semigroup property):
\begin{equation}
[\mathcal{L}_{\beta}\phi](x)=\int_X{\rmd y \, \,\rme^{\beta (f(y)-y)}\delta(f(y)-x)\phi(y)}
\end{equation}
The key idea is that if we are able to derive an analytic expression for the generating function, we can then extract all transport moments:
\begin{equation}
\langle(x_t-x_0)^q \rangle= \left.\left(\frac{\partial^q}{\partial \beta^q}   G_t(\beta)\right) \right|_{\beta=0}
\end{equation}
The generating function may be expressed as
\beq
  G_t(\beta)=\int{\rmd x [\mathcal{L}^t_{\beta}\rho_{\rm in}](x)}
\eeq
where $\rho_{{\rm in}}$ is the density of initial conditions: the asymptotic growth of $G_t(\beta)$ in this way is dominated by the leading eigenvalue of the generalized transfer operator, as in equilibrium statistical mechanics of lattice systems, namely
\beq
  G_t(\beta) \sim \lambda^t(\beta)
\eeq
and then
\begin{equation}\label{der-eigen}
\langle (x_t-x_0)^q \rangle \sim \left. \left(\frac{\partial^q}{\partial \beta^q}  \lambda^t (\beta)\right)\right|_{\beta=0}
\end{equation}
The main problem thus consists in computing the leading eigenvalue (and its derivatives): before turning to this point we have to describe the particular class of dynamical systems we want to study: generally we are interested in spatially extended periodic  systems where the dynamics over the unbounded phase space is generated by replicas of a map on a finite fundamental domain (like the Lorentz gas on a square lattice, where the dynamics on a fundamental cell corresponds to the Sinai billiard). From now on we illustrate  the case of a one-dimensional  map on the real
line, even though the method is by no means limited to this context. The key property
of the dynamical system under investigation is represented by the symmetry properties
\begin{eqnarray}
f(-x) & \, = \,  & -f(x) \label{nodrift}\\
f(x+n) & \, = \, & n+f(x) \label{cells}
\label{symm-P}
\end{eqnarray}
for any $n \in \mathbb{Z}$. The property (\ref{nodrift}) guarantees the absence of a net drift
while with  (\ref{symm-P}) we impose that the map on the real line is obtained by lifting a circle
map $\hat{f}(\theta)$ defined on the unit torus as
\begin{equation}
\hat{f}(\theta)\,=\, \left. f(\theta)\right|_{{\rm mod}\,\,1} \qquad \theta \in {\mathcal{T}}=[0,1)
\label{tor-f}
\end{equation}
We may thus split the $f$ evolution into a box integer plus a fractional part:
$x_n=N_n+\theta_n$,
where
\begin{eqnarray}
N_{n+1} &\,=\, & N_{n} + \sigma(\theta_n) \label{intf}\\
\theta_{n+1}&\, =\,&   \hat{f}(\theta_n) \label{fracf}
\end{eqnarray}
where $\sigma(\theta)=[f(\theta)]$ is the function returning the integer part of the argument.

When considering the generating function  (\ref{genfunc}) translation invariance  of the system permits to restrict the average over the elementary cell. The transfer operator  on the fundamental cell is written  as:
\begin{equation}
[\mathcal{L}_{\beta} \rho \,](\phi)=\int_{\mathcal T} \, \rmd \theta \rme^{\beta (\hat{f}(\theta)-\theta+\sigma(\theta))} \delta(\hat{f}(\theta)-\phi) \rho(\theta)
\end{equation}
where all quantities refer to the torus map once we attach to every point in ${\mathcal T}$ the correspondent jumping number $\sigma(x)=[f(x)]$ \cite{PD1,PD2}.

The leading eigenvalue is the inverse of the smallest $z$ solving the secular equation:
\begin{equation}
\det(1-z\mathcal{L}_{\beta})=F_{\beta}(z)=0
\end{equation}
($F_{\beta}(z)$ is usually called the spectral determinant).
Using the formal relationship:
\begin{equation}
\det (1-z\mathcal{L}_{\beta})= \exp \Tr \ln (1-z\mathcal{L}_{\beta})=\exp \left(-\sum_{n=1}^{\infty}\frac{z^n}{n}\Tr \mathcal{L}^n_{\beta}\right)
\end{equation}
we are lead to computing the formal trace of $\mathcal{L}^n_{\beta}$:
\begin{eqnarray}
\Tr \mathcal{L}^n_{\beta}&=&\int \rmd x  \,\,\delta(\hat{f}^n(x)-x)\rme^{(\hat{f}^n(x)-x+\sigma_n(x))}=\\
&=&\sum_{x=\hat{f}^n(x)} \frac{ \rme^{(\beta \sigma_n(x))}}{|1-\Lambda_n(x)|} \label{somma}
\end{eqnarray}
where
\begin{eqnarray}
\Lambda_n(x)&=&\prod_{i=0}^{n-1}{\hat{f}'(\hat{f}^i(x))} \\
\sigma_n(x) &=&\sum_{j=1}^n \sigma(f^j(x))
\end{eqnarray}

and the sum runs over all periodic orbits of the system under study.

We may now proceed to derive the standard expression for spectral determinants and zeta functions: the first step consists in using a geometric series expansion for the denominator in (\ref{somma}):
\begin{equation}
\Tr \mathcal{L}^n _{\beta} =\sum_{k=0}^{\infty} \sum_{x=\hat{f}^n(x)} \frac{ \rme^{(\beta \sigma_n(x))}}{|\Lambda_n(x)|(\Lambda_n(x))^k}
\end{equation}
We remark that the geometric series expansion makes sense only if we deal with a hyperbolic system, for which $|\Lambda_n(x)|>1$ for any point and iteration order (differently from what happens for intermittent systems). Next we perform a subtle step: the sum is over periodic points of any possible order $n$: so for instance fixed points appear at any order, and every $p$ orbit reappears whenever $p$ divides $n$. Moreover all points of a given periodic orbit contribute to the trace, but they all yield the same weight, as instabilities and jumping numbers $\sigma_n(x)$  are the same for every points of the same periodic orbit. So we may convert the sum over periodic points to a sum over \emph{orbits} $p$, each taken once: we furher denote by $\Lambda_p$ and $\sigma_p$ respectively the instabilities and jumping numbers of the orbit $p$, each computed with respect to its prime period $n_p$. In this way we get:
\begin{equation}
\det (1-z\mathcal{L}_{\beta})=\exp \left(-\sum_{\{p\}}\sum_{r=1}^{\infty}\sum_{k=0}^{\infty} \frac{z^{r \cdot n_p}}r \frac{\rme^{\beta \sigma_p \cdot r}}{|\Lambda_p|^r \Lambda_p^{k\cdot r}} \right)
\end{equation}
where $\{p\}$ denotes the set of periodic orbits, each taken once, $n_p$ being the prime period of the orbit. The sum over $r$ reproduces each repetition of a single orbit $p$ in the original summation over periodic points: this sum may be performed exactly, leading to:
\begin{eqnarray}
\det (1-z\mathcal{L}_{\beta})&=&\exp\left(\sum_{\{p\}}\sum_{k=0}^{\infty} \ln\left( 1-\frac{z^{ n_p} \rme^{\beta \sigma_p}}{|\Lambda_p| \Lambda_p^{k}} \right) \right)=\\
&=&\prod_{k=0}^{\infty} \prod_{\{p\}}\left( 1-\frac{z^{ n_p} \rme^{\beta \sigma_p}}{|\Lambda_p| \Lambda_p^{k}} \right)
\end{eqnarray}
Thus the spectral determinant may be expressed by an infinite product over the so called generalized zeta functions:
\begin{eqnarray}
\zeta^{-1}_{(k)\beta}(z) &= &\prod_{\{p\}}\left( 1-\frac{z^{ n_p} \rme^{\beta \sigma_p}}{|\Lambda_p| \Lambda_p^{k}} \right)\\
F_{\beta}(z)&=&\prod_{k=0}^{\infty} \zeta^{-1}_{(k)\beta}
\end{eqnarray}
and the leading zero can be calculated from the zero order zeta, the \emph{dynamical zeta function} $\zeta^{-1}_{(0)\beta}$.

We stress again that the quantities that enter the definition of zeta functions are the prime period $n_p$ of the orbit $p$, its
instability $\Lambda_p=\prod_{i=0}^{n_p-1}\hat{f}'(\hat{f}^i(x_p)$ and the integer factor $\sigma_p$,
that accounts for the orbit's behaviour once we unfold it on the real line. As a matter of fact, given
any point $x_p$ belonging to $p$ ({\em i.e.} $\hat{f}^{n_p}(x_p)=x_p$) we may either have that it is
also a periodic point of the lift $f$ ($f^{n_p}(x_p)=x_p$), or it might be a running mode,
{\em i.e.} $f^{n_p}(x_p)=x_p+\sigma_p$, with $\sigma_p \in \mathbb{Z^*}$ (see (\ref{tor-f})).

It is well known that zeta functions calculation may be recast in terms of perturbative expansions (the order being determined by the longest prime period available), that yield exponentially good estimates \cite{POE}.
In order to show that let's use a simplified example: suppose there is a complete grammar in the simple alphabet $\{0,1\}$, calling for shortness $t_p=\frac{z^{ n_p} \rme^{\beta \sigma_p}}{|\Lambda_p| \Lambda_p^{k}}$ we can expand the infinite product in series:
\begin{eqnarray}
\zeta^{-1}_{(k)\beta}(z)&=&\prod _{\{p\}}(1-t_p)=\\
&=&(1-t_0)\cdot(1-t_1)\cdot(1-t_{01})\cdot(1-t_{001})\cdot(1-t_{011})\cdots\nonumber\\
&=& 1-t_0 -t_1 +\nonumber\\
 &&- (t_{01}-t_0t_1)-(t_{001}-t_0t_{01})-(t_{110}-t_1t_{01})-\cdots\label{cyclexp}
\end{eqnarray}
We can notice that there are  \emph{fundamental terms} $t_0$ and $t_1$  while others \emph{curvature term} are made of a prime cycle shadowed by combinations of shorter cycles. Completely hyperbolic systems with a finite grammar yield curvature corrections that are exponentially small with the maximum period, so that (\ref{cyclexp}) (power law expansion of the zeta function) is indeed a perturbative series.

We now sketch the technique we will actually use to investigate arbitrary order transport moments: first we reorder the zeros $\bar{z}_i$ of the spectral determinants so that the dominant ones come first and  write in general:
\begin{equation}
F_{\beta}(\rme^{-s})=\prod_{i}(1-\bar{z}_i \rme^{-s})
\end{equation}
so that
\begin{equation}
\frac{d}{d s}\ln  F_{\beta}(\rme^{-s})=\sum_{i}\frac{\bar{z}_i \rme^{-s}}{(1-\bar{z}_i \rme^{-s})}
\end{equation}
and if we now take the inverse Laplace transform, we get
\begin{equation}\label{eqnLaplace}
\frac{1}{2 \pi \rmi}\int_{a - \rmi \infty}^{a+ \rmi \infty} \, ds\, \rme^{sn}
\, \frac{d}{ds} \ln \left[  F_{\beta}(\rme^{-s}) \right]=\sum_i \bar{z}_i^n
\end{equation}

As the  asymptotic behaviour is dominated by the leading eigenvalue we may extract it by using the dynamical zeta function instead of the spectral determinant, thus
\begin{equation}
 \bar{z}_0^n(\beta) \sim \frac{1}{2 \pi \rmi}\int_{a - \rmi \infty}^{a+ \rmi \infty} \, ds\, \rme^{sn}
\, \frac{d}{ds} \ln \left[ \zeta_{(0)\beta}^{-1}(\rme^{-s}) \right]
\end{equation}
 Remember that by probability conservation we have  $\bar{z}_0(0)=1$ and all  transport properties are captured by small $\beta$ expansion (\ref{der-eigen}):
\begin{eqnarray}\label{gen-mom}
\sigma_k(n)&\,=\,& \langle (x_n -x_0 )^k \rangle_0 \,=\, \left. \frac{\partial^k \,} {\partial \beta^k}
G_n(\beta) \right|_{\beta=0} \nonumber \\
 &\,\sim\,& \left. \frac{\partial^k \,} {\partial \beta^k} \frac{1}{2 \pi \rmi}\int_{a - \rmi \infty}^{a+ \rmi \infty} \, ds\, \rme^{sn}
\, \frac{d}{ds} \ln \left[ \zeta_{(0)\beta}^{-1}(\rme^{-s}) \right] \right|_{\beta=0}
\end{eqnarray}
The evaluation of the integral on the right hand side of (\ref{gen-mom}) requires dealing with
high order derivatives of a composite function: this is generally accomplished by making use of Fa\`a di Bruno formula:
\begin{eqnarray}
\frac{d^n \,}{dt^n}H(L(t))= \sum_{k=1}^n\,\sum_{k_1,\dots ,k_n} \,
\frac{n!}{k_1! \cdots k_n!}\frac{d^k H}{dL^k}  (L(t))\cdot
B_{\bi{k}}(L(t))
\label{fdb}\\
B_{\bi{k}}(L(t))=\left(\frac{1}{1!}\frac{dL}{dt}\right)^{k_1} \cdots \left(\frac{1}{n!}\frac{d^nL}{dt^n}\right)^{k_n} \, \, \, \,  \quad \, ~\\
\bi{k}=\{k_1,\dots k_n\} \,\hbox{\rm with}\, \sum k_i=k ,\quad \sum i\cdot k_i=n
\end{eqnarray}

Remember that the characterization of anomalous transport is
captured by the function $\nu(q)$ (the fundamental quantity we
want to evaluate) that expresses the asymptotic growth of moments
of arbitrary order:
\begin{equation}
\langle | x_t-x_0 |^q \rangle \, \sim \, t^{ \nu(q)} \,
\label{nu-mom}
\end{equation}
We already mentioned how ordinary diffusion yields  $\nu(q)=q/2$, while anomalous transport typically shows a non trivial behaviour, which cannot
be encoded by a single exponent, but rather typically exhibits a
phase transition \cite{Ark,V-AD,OTT}.
In next  sections
we will apply the formalism to both normal and anomalous examples. These computations accomplish a twofold goal: besides checking the
theory, they will also illustrate how subtle features of the underlying
dynamical system are automatically included in the formalism.
\section{Normal transport}
We now show how the technique is able to reproduce the well known exponents $q/2$ for  the transport moments of a completly chaotic dynamic. This oversimplified example is a warmup exercise before facing more involved (and interesting) cases.

 Consider the following map on the real line
\begin{equation}\label{simplemap}
f(x)=\left\{ \begin{array}{ll}
  ax & x \in[0,1/2]\\
  a(x-1)+1 & x \in (1/2,1]
\end{array} \right.
\end{equation}
For a generic value of the parameter a symbolic dynamics of (\ref{simplemap}) is quite complicated (and this leads to interesting consequences, like fractal dependence of the diffusion constant on the slope $a$ \cite{KL}): on the contrary by choosing an integer slope the torus map consists of complete branches: for example for   $a=2m$, $m\in \mathbb{N}$ the corresponding torus map consists of $2m$ complete branches and the grammar is thus unrestricted in a $2m$ alphabet. Moreover the linearity of the map leads to complete cancellations of the curvature terms: whenever the weights associated to each cycle depend multiplicatively on the symbolic code entries, the contribution of a cycle and corresponding pseudocycles cancel out exactly and zeta functions pick up contribution only from fixed points.

In this case there are $2m$ fixed points: the instabilities are trivially $\Lambda_i=a$ for every fixed point while the jumping numbers associated to each branch are clearly:
$0,1,\cdots,(a-1),-(a-1),\cdots,-1,0$.

 The zeta function for the torus map  is then:
\begin{equation}
\zz = 1-\frac{2z}{a}\left(1+\sum_{k=1}^{a/2-1}\cosh(\beta k)\right)
\end {equation}

Long time behaviour of the q-th moment  is captured by (\ref{gen-mom}). Following \cite{POE} we concentrate first on $\beta$ derivatives
\beq
\left.\der{n}{\beta}{}\ln \zz\right|_{\beta=0}
\eeq

Fa\`a di Bruno's expansion (\ref{fdb}) generates:
\beq
\sum_{k=1}^n\,\sum_{k_1,\cdots k_n}
\, \left. \frac{1}{\zz^k}  \cdot B_{\bi{k}}(z)\right|_{\beta=0}
\label{normal-fdb}
\eeq
and a generic term contributing to the product in $B_{\bi{k}}(z)$ is now
\beq
\der{k_j}{\beta}{}\zz_{|\beta=0} = \left\{ \begin{array}{ll}
  \frac{2z}{a}\sum_{l=1}^{a/2-1}{l^{k_j}}& \quad k_j \,\, \hbox{\rm even}\\
 0& \quad k_j \, \,\hbox{\rm odd}
 \end{array} \right.
\eeq
Each term in (\ref{normal-fdb}) has a different asymptotic behaviour and contribute differently to the transport moment once we perform the derivative with respect to $s$ and take the inverse Laplace transform in (\ref{gen-mom}).  The long time behaviour of the q-th transport moment is  captured by choosing the leading singularity of the $\beta$ derivatives near $z=1 \, \, (s=0)$.

 The numerator in (\ref{normal-fdb}) doesn't diverge for $z\to 1$ and the divergence comes only from the denominator:
\beq
\frac{1}{\zz^k}\ca (1-z)^{-k}
\eeq
The leading singularity  comes from the term with the largest possible choice for $k$ and hence all $k_j =2 $ (remember that for $k_j$ odd the numerator vanish). This leads to the estimate
\begin{equation}
\der{n}{\beta}{}\ln \zz_{|\beta=0}     \sim (1-z)^{-n/2}
\end {equation}
and, performing the inverse Laplace transform, this yields
$\nu(q)=q/2$, typical of normally diffusing systems (where one is
then interested in obtaining estimates of the prefactor, namely
the diffusion constant). We turn now on more interesting case of
intermittent diffusion where the presence of marginal points
deeply modified the structure of transport exponents.
\section{Intermittent systems}
 In the last few years it has been realized \cite{gg,ACT,per,pr,isola} that weakly chaotic systems (in particular
one dimensional intermittent maps, or infinite horizon Lorentz gas models lead to more
complicated analytic structure of  zeta functions, which typically exhibit branch
points. In view of the inverse Laplace formula (\ref{gen-mom}) the modified analytic structure
of the zeta function may induce anomalous behaviour (nonlinear diffusion \cite{ACL,POE}).

\begin{figure}[h!]
\centerline{\epsfxsize=12.cm \epsfbox{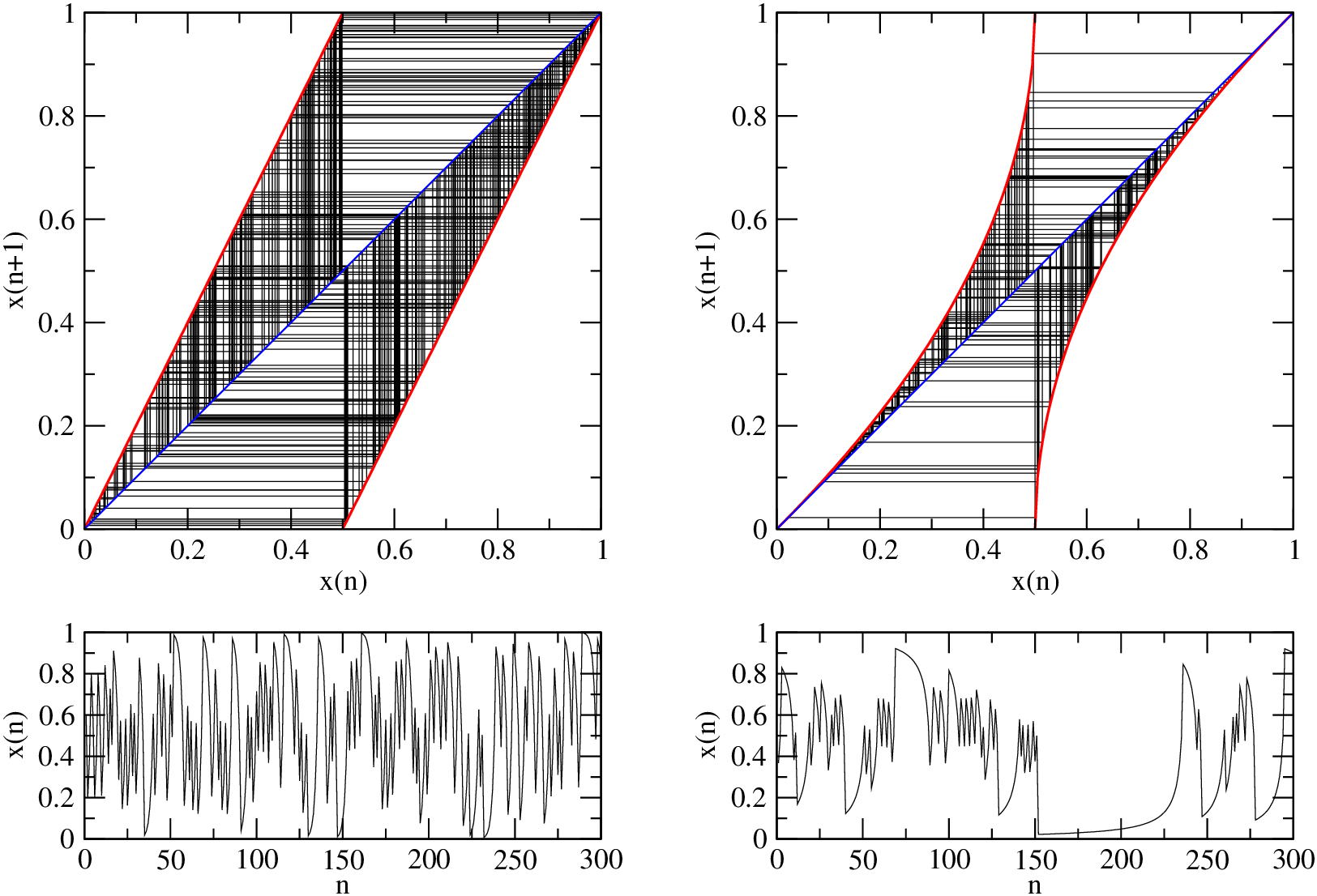}} 
\vskip4mm
\caption{\label{normvsint}A segment of trajectory ($300$
iterations) with the same initial condition is showed for two
torus maps: a completely hyperbolic  map (left)  and one with
intermittent fixed points (right).}
\end{figure}

The qualitative  difference between an hyperbolic system and an intermittent one can be well appreciated in fig. (\ref{normvsint}) where a long segment of an  orbit for both cases is plotted against time.

 In the intermittent case chaotic wandering is interrupted by intervals of regular motion, generated by points that start their motion near the marginal point and slowly move away from it: we will see that there is a power low separation for nearby trajectories instead of the exponential growth we normally expect. In the next subsection we will use the thermodynamic formalism in order to detect the presence of power law behaviour in instabilities spectrum while in the following subsections we explain how it is possible to implement cycle expansion also in this anomalous case.

\subsection{Thermodynamic formalism}
 In order to obtain information about the scaling behaviour of periodic orbits instabilities we turn our attention to a thermodynamic analysis of this set.
Consider \emph{all} periodic  points of $n$ order  $x_{i(n)}$,
$i=1 \dots N_n$. Consider now the partition of the unit interval
with the intervals: \beq I_{i(n)}=(x_{i(n)},x_{i+1(n)}) \qquad i=1
\dots N_n \eeq whose length we denote by $l_{i(n)}=|I_{i(n)}|$,
and construct with them the following partition function: \beq
Z_n(\tau)=\sum_{i=1}^{N_n-1}{(l_{i(n)})^{-\tau}} \eeq where the
parameter $\tau$ is the formal analogue of an inverse temperature.
By varying $\tau$ one probes for all  the relative different
scaling of the intervals in the partition function sum
\cite{THER}. For example for $\tau \to +\infty$ the partition
function will be dominated by the thinnest intervals while for
$\tau \to -\infty$ the fattest ones prevail.

Scaling properties may be analyzed \cite{ACK} by writing
\beq
l_{i(n)}=(N_n)^{-\mu_{i(n)}}
\eeq
where we have introduced the local scaling indices $\mu_{i(n)}$: it is important to notice that they will be finite if the interval shrink exponentially with $n$ (as typically $N_n \sim a^n$) while will go to zero as $(\ln n)/n$ for a power law behaviour.

The (asymptotic) free energy is then defined as
\ba
g(\tau)&=&\lim_{n \to \infty}{\frac1{\ln(N_n)} \ln{Z_n}(\tau)}\\
&=&\lim_{n \to \infty}{\frac1{\ln(N_n)} \ln{\sum_i{N_n^{\tau
\,\mu_{i(n)}}}}} \ea
 from which we can extract the average scaling index:
\beq \mu(\tau)=g'(\tau)=\lim_{n \to \infty}
\frac{\sum_i{\mu_{i(n)}N_n^{\tau \,\mu_{i(n)}}}}{\sum_i N_n^{\tau
\mu_{i(n)}}} \eeq

\begin{figure}[h!]
\centerline{\epsfxsize=10.cm  \epsfbox{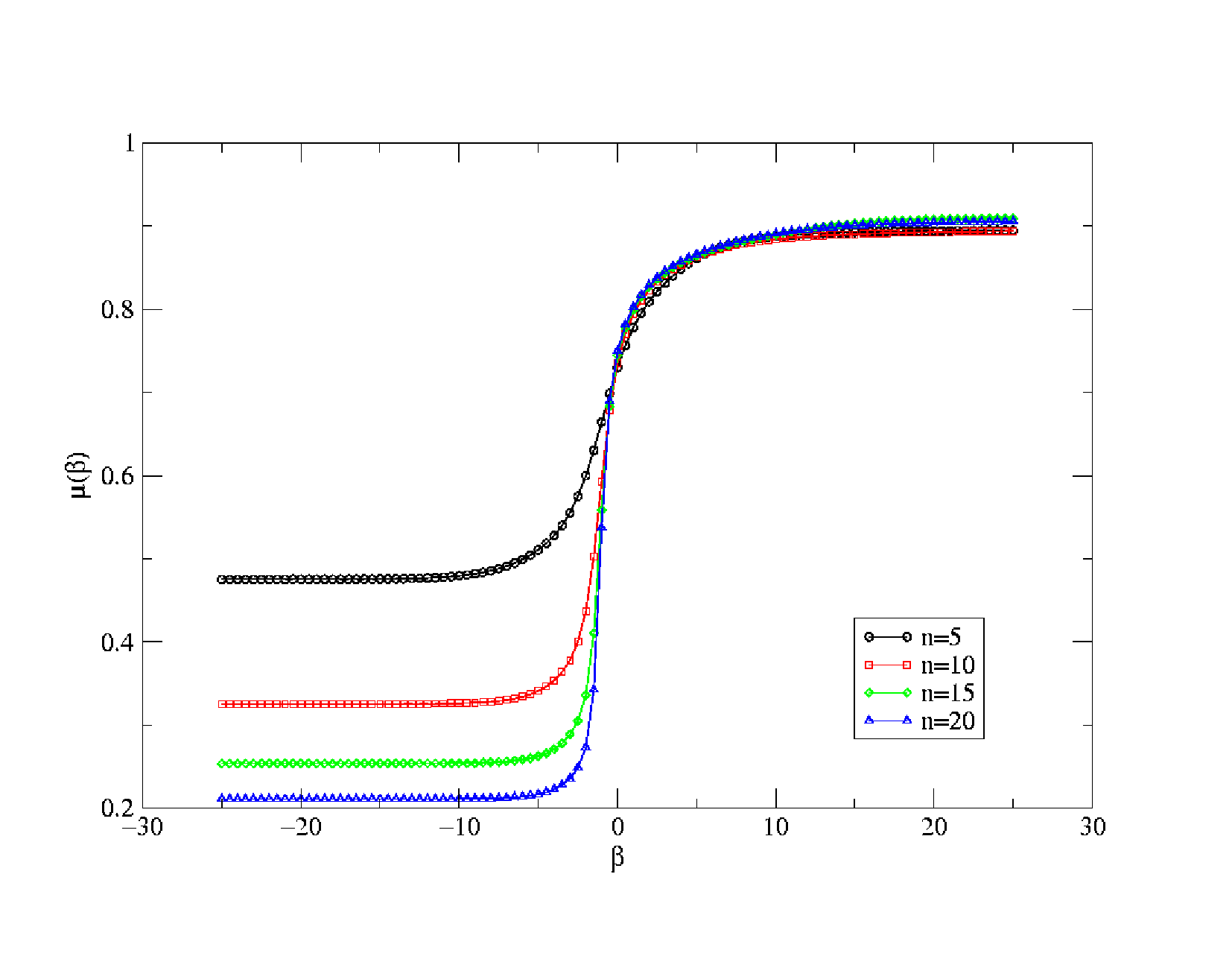}} \vskip4mm
\caption{\label{f3} Approximated scaling index functions
$\mu_n(\tau)$ versus $\tau$ for the map in (\ref{arkmap}) with
$\gamma=2.5$ and with n=5,10,15,20.}
\end{figure}

Figure (\ref{f3}) shows four  approximated scaling index functions
$\mu_n(\tau)$  calculated for an intermittent map: this map  has
two complete branches and thus the number of periodic points of
order $n$ grows as $2^n$ while the presence of marginal points
generates a family of intervals that shrink polynomially with $n$
(the exact form of this map will be presented as second example in
following subsection (see (\ref{arkmap})). We can note that by
increasing the hierarchical order the scaling index  converge to
finite values for $\tau >0$ while it slowly goes to zero for $\tau
<0$, as we expect due to intervals that shrink with power law
behaviour.

The thermodynamic analysis may be pushed further by rewriting the
partition function as: \beq\label{zintegral}
Z_n(\tau)=\int_{\mu_{min}}^{\mu_{max}}{\rmd \mu
N_n^{(s_n(\mu)+\mu\tau)}} \eeq where   $N_n^{s_n(\mu)}$ is the
density of intervals with scaling index $\mu$. As the number of
periodic orbits increases the integral can be approximated by the
saddle point method: \beq N_n^{g_n(\tau)}=C(n) \Big[-\frac{2 \pi
\ln N_n}{S''_n(\tau)}\Big]^{\frac12} N_n^{(S_n(\bar{\mu})+\tau
\bar{\mu})} \eeq where $\bar{\mu}$ is the solution of the extremum
condition \beq \left.\frac{dS_n(\tau)}{\rmd
\mu}\right|_{\bar{\mu}}=-\tau \eeq and $g_n(\tau)$ is the n-th
order approximation to the free energy.

The scaling spectrum $s_n(\mu)$
  is a highly irregular function of $\mu$ for finite $n$: in practise one evaluates $g_n(\tau)$ and replaces $s(\mu)=\lim_{n \to \infty} s_n(\mu)$ by its convex envelope $S(\mu)$, evaluated in the thermodynamic limit saddle point evaluation of (\ref{zintegral}).

\begin{figure}[h!]
\centerline{\epsfxsize=10.cm  \epsfbox{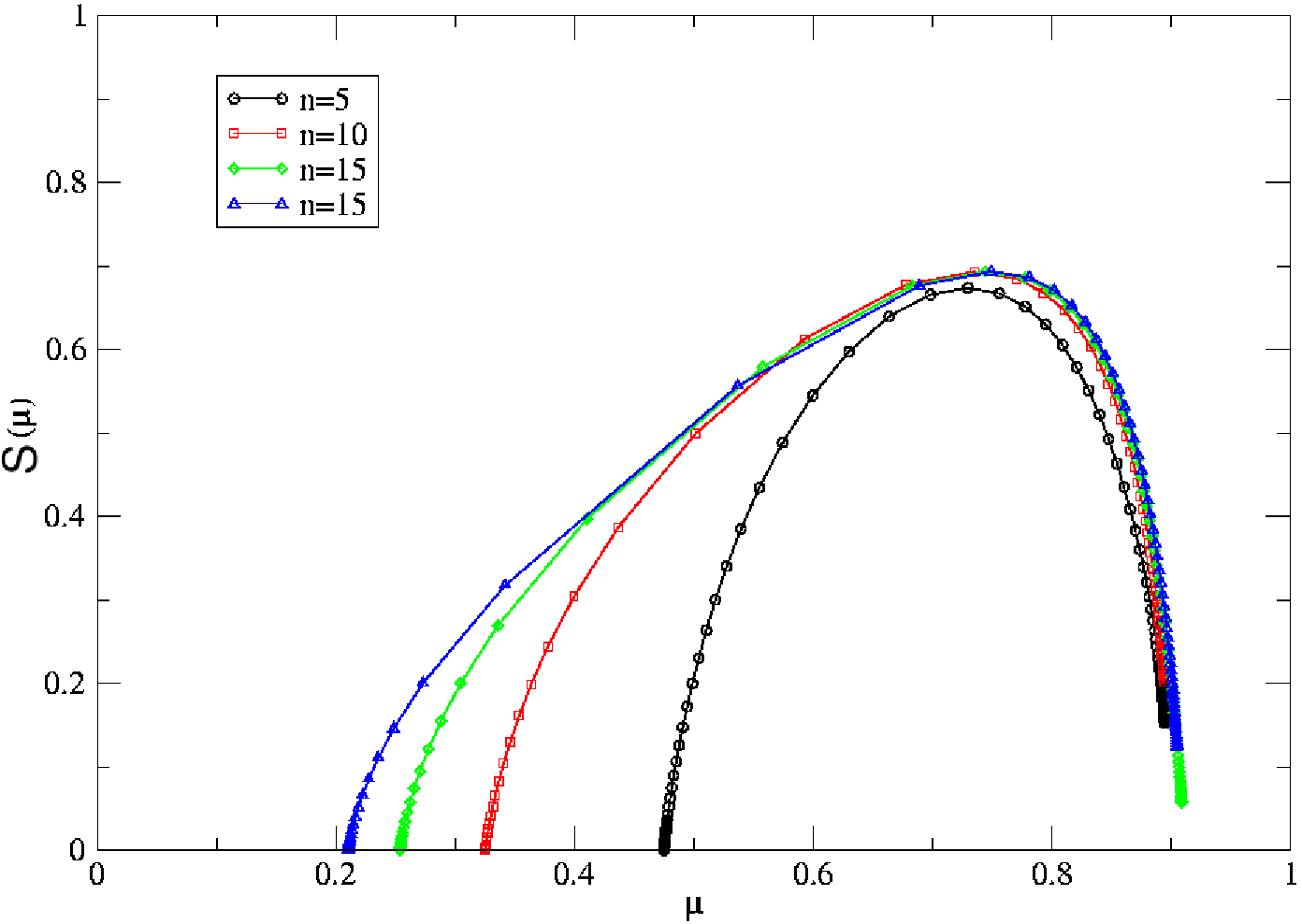}}
\vskip4mm
\caption{\label{f4} Approximated  entropies $S_n(\mu)$ versus $\mu$ for the map (\ref{arkmap}) with $\gamma=2.5$ and with n=5,10,15,20.}
\end{figure}

For the very same example used in calculating the scaling index of figure (\ref{f3}) we now compute the convex envelope entropy $S_n(\mu)$. While for a complete hyperbolic dynamic we expect a convergence to a curve with strictly positive support, in our case  marginality  yields  a tail in the small scaling region of $S(\mu)$, which slowly advances towards zero as the hierarchical order increases, see figure (\ref{f4}): the right branch of entropy is stable and it is representative of exponential scalings while the leftmost branch slowly moves towards zero, being ruled by intervals shrinking with power laws.

\subsection{Transport}
We now apply  the technique to two classes of one dimensional maps, where deviations from
fully chaotic behaviour are provided by marginal fixed points of intermittent type \cite{PM}. The prototype example of this class of systems is provided by the mapping  on the unit interval
\beq\label{Pomeau}
\left.x_{n+1} \ca x_n + x_n^{\gamma}\right|_{{\rm mod}\,\,1}
\eeq
where $\gamma >1 $ is called the intermittency exponent.

The presence of the marginal fixed point causes problems in using zeta function techniques in a straightforward manner: this fixed point has $|\Lambda|=1$, which leads to divergences (see (\ref{somma})). The way out of this problem is to prune the $\bar{0}$ fixed point away: from a symbolic dynamics point of view this leads to an infinite alphabet, since all $0^n1$ fixed points lack curvature counterterms: the fundamental cycle part  of the zeta function thus becomes (cfr. (\ref{cyclexp}))
\beq\label{intermcycle}
\zeta^{-1}_{\rm fund}(z)=1-t_1-t_{01}-t_{0^21}- \cdots - t_{0^n1}- \cdots
\eeq
so even if curvature corrections are neglected we do not have a finite order polynomial: moreover the inclusion of curvature correction even in simple cases is by no means trivial \cite{ACT}. As we will see by considering specific examples, the analytic structure of (\ref{intermcycle}) will be strongly  dependent on the intermittency exponent, through the behaviour  of stability of cycles $0^n1$ coming closer and closer to  the marginal fixed point.
\subsection{Intermittent diffusion I}
We can turn now on a specific example. The symmetry requirements are  the same as the completely chaotic  example (see relations (\ref{nodrift}) and (\ref{cells})).
\begin{figure}[t]
\centerline{\epsfxsize=6.5cm \epsfbox{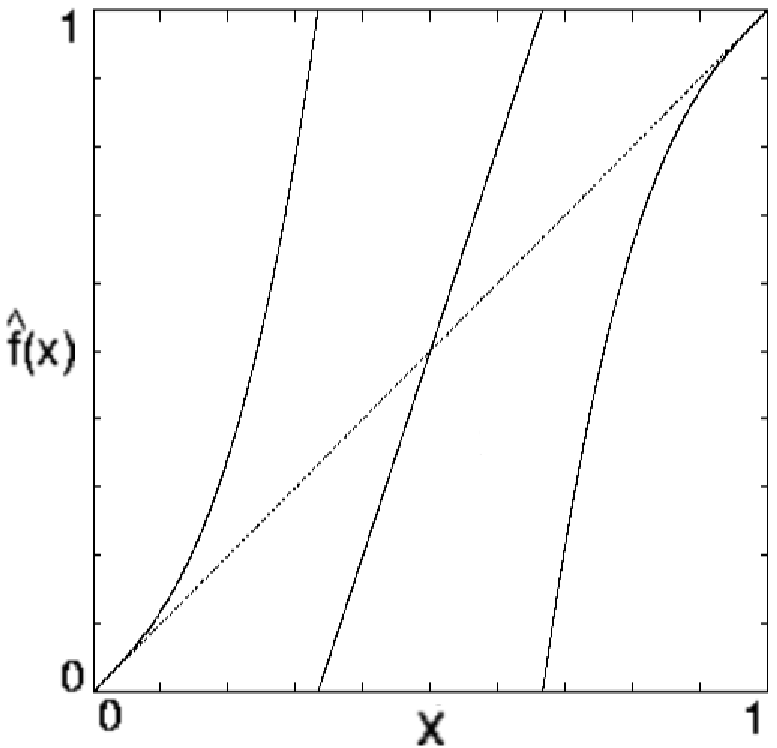}
\hspace{2mm}\epsfysize=6.5cm \epsfbox{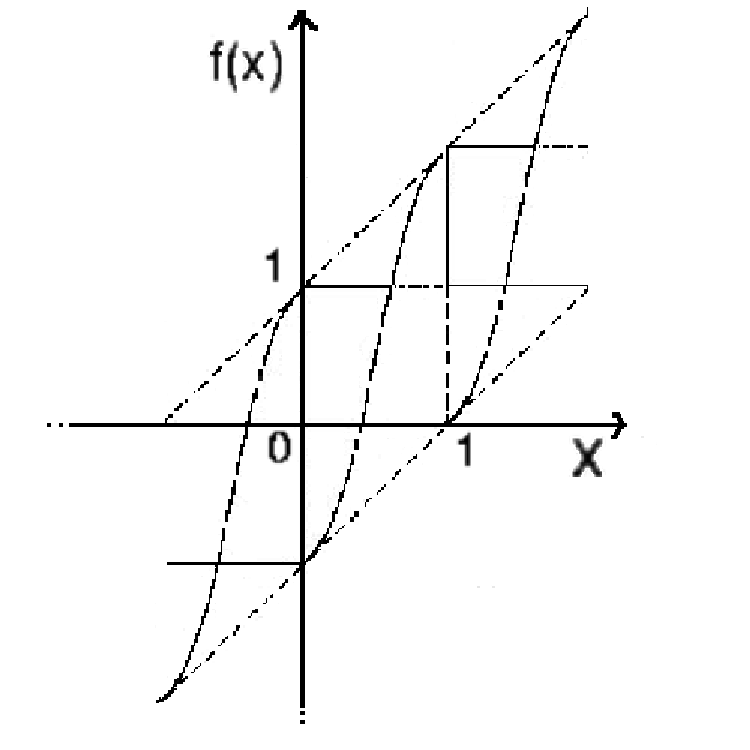}}
\vskip4mm
\caption{\label{f5}A torus map with intermittent fixed points, and the corresponding lift.}
\end{figure}

The map on the fundamental cell (see figure (\ref{f5}))  consists of two intermittent branches and one central hyperbolic region. If we assign jumping numbers  $\sigma_L=-1$ to the left branch, $\sigma_{C}=0$ to the central one and
$\sigma_{R}=+1$ to the right branch we get the lift map of figure (\ref{f5}).
Whenever a particle remain trapped near a marginal point the corresponding unfolded trajectory on the real line consists of successive jumps on the neighbouring cell.

The presence of complete  branches leads to an unrestricted grammar in the symbolic code $\{L,C,R\}$, where the corresponding partition obviously consist of the inverse image of the unit interval under the corresponding three branches functions $f_L, f_C, f_R$. However we have to prune marginal fixed points $\bar{L}$ and $\bar{R}$ away, as explained before: the grammar is thus unrestricted on a  \emph{countable} new alphabet:
\beq
\{ C,L^iC,L^jR,R^kC,R^lL\} \quad i,j,k,l \in \mathbb{N}_+
\eeq
The boundary points of such a partition can be constructed by sequences of iterates of the inverse map (L or R branches):
$f^{-1}_L(1),f^{-2}_L(1),\dots, f^{-1}_R(0),f^{-2}_R(0),\dots$. The corresponding zeta function will thus consist of a (non polynomial) fundamental part (\ref{intermcycle}), together with curvature corrections: by a straightforward generalization of Gaspard-Wang piecewise linear approximation \cite{plapp} of Pomeau-Manneville map, we can obtain an approximate model where all curvature terms
vanish and the zeta function can be explicitly written without losing important dynamical features \cite{isolapl}.  The power law scaling of the stability of cycles coming closer and closer to  the marginal points can be extimated as
\beq
\Lambda_{i} \sim i^{-(\alpha +1)}
\eeq
where $i$ indicates the number of iteration the orbits needs to escape from the laminar phase, and
\begin{equation}\label{stab-z}
\alpha=1/(\gamma-1)
\end{equation}
$\gamma$ being the intermittency exponent of the L-R branches (see (\ref{Pomeau})).

The zeta function is:
\begin{equation}
\label{pl-dz}
\zeta_{(0)\beta}^{-1}(z)\,=\, 1-az-bz\sum_{k=1}^{\infty}\, \frac{z^k}{k^{\alpha+1}}\cosh (\beta k)
\end{equation}
where $a$ and $b$ are fixed by specifying the central region slope and the normalization
condition $\zeta_{(0)0}^{-1}(1)=0$.

When equipped with information about the analytic structure of the zeta function, from (\ref{fdb})
we may single out the leading singularity in the logarithmic derivative of the zeta function, and then
estimate the asymptotic behaviour of (\ref{gen-mom}).
In our previous example the only divergence for $ z\to1$ in derivatives expansions (\ref{fdb}) comes from the denominator while here
the appearance of the Bose function (see (\ref{pl-dz}))
\begin{equation}
\label{Bf}
g_{\mu}(z)\,=\,\sum_{l=1}^{\infty}\, \frac{z^l}{l^{\mu}}
\end{equation}
indeed may alter the analytic features. We recall how the behaviour as $z \to 1^{-}$ depends upon $\mu$ (related to the intermittency  exponent)
\begin{eqnarray}
g_{\mu}(z) \sim  \left\{
\begin{array}{ll}
(1-z)^{\mu -1} \quad & \mu<1 \\
\ln (1-z) \quad & \mu=1 \\
\zeta(\mu)+C_{\mu}(1-z)^{\mu-1}+D_{\mu}(1-z) \quad & \mu \in (1,2) \\
\zeta(2)+C_2(1-z) \ln (1-z) \quad &  \mu=2 \\
\zeta(\mu)+C_{\mu}(1-z) \quad & \mu> 2
\end{array} \right.
\label{B-as}
\end{eqnarray}

We stress again that this functions appear as a consequence of particular sequences of orbits coming closer and closer to the marginal fixed points whose stability increases only polynomially with the period \cite{plapp,POE}, a clear signature
of local deviation from typical hyperbolic behaviour (which is ruled by exponential instability
growth).

Now look at a generic term in (\ref{normal-fdb}) and denote it by ${\cal D}_{k_1 \dots k_n}$. Taking into account that
\begin{eqnarray}
\left. \frac{\partial^i \, }{\partial \beta^i}\zeta_{(0)\beta}^{-1}(z)\right|_{\beta=0}\,\sim \,
\left\{
\begin{array}{ll}
0 \qquad & i \,\,\mathrm{odd} \\
zg_{\alpha+1-i}(z) \quad & i\,\, \mathrm{even}
\end{array}
\right.
\label{djz}
\end{eqnarray}
 we have that
\begin{equation}
\label{typ-fdb}
{\cal D}_{k_1 \dots k_n}\,\sim\, \frac{1}{(\zeta_{(0)0}^{-1}(z))^k}\prod_j\,(g_{\alpha+1-j}(z))^{k_j}
\,=\,\frac{{\cal D}^{+}_{k_1 \dots k_n}}{{\cal D}^{-}_{k_1 \dots k_n}}
\end{equation}
where the ${\cal D}^{+}$ picks up the contributions from the product of Bose functions, and
all $j$ must be even, due to (\ref{djz}).
First we consider the case $\alpha \in (0,1)$, which corresponds to intermittency exponent
$\gamma > 2$: we have ${\cal D}^{-}_{k_1 \dots k_n}\sim (1-z)^{k\alpha}$, that,
together with (\ref{B-as}), implies that the dominant singularity is of the form
\begin{equation}
\label{ddom}
{{\cal D}_n}\,\sim \, \frac{1}{(1-z)^{\rho}}
\end{equation}
where $\rho$ is determined by
\begin{equation}
\label{rho}
\rho=  \sup_{\{k_1 \dots k_n\}} ( k\alpha + \sum_j (j-\alpha)k_j) =n
\end{equation}
Once plugged into (\ref{gen-mom}) this leads to the estimate
\begin{equation}
\label{nu-norm}
\nu(q)\,=\, q
\end{equation}
which means that the whole set of transport moments is ruled by ballistic behaviour (at least
for even exponents, where the method applies).
We now turn to the more subtle case $ \alpha > 1$: since the dynamical zeta function has a
simple zero we get ${\cal D}^{-}_{k_1 \dots k_n}\sim (1-z)^k$, while the terms appearing in
${\cal D}^{+}$ modify the singular behaviour near $z=1$ only for sufficiently high $j$
\begin{equation}
g_{\alpha+1-j}(z)\,\sim\,
\left\{
\begin{array}{ll}
(1-z)^{\alpha -j}\quad & j> \alpha \\
\zeta(\alpha +1-j) \quad & j< \alpha
\end{array}
\right.
\label{ggg}
\end{equation}
If all $\{j\}$ are less than $\alpha$ then the singularity is determined by ${\cal D}^{-}$: keeping in
mind that the highest $k$ value is achieved by choosing $j=2$ and $k_2=n/2$, we get, by
proceeding as before
\begin{equation}
\label{mom-}
\nu(q)\,=\,\frac{q}{2} \qquad q < \alpha
\end{equation}
When $q$ exceeds $\alpha$ we have to take into account possible additional singularities in
${\cal D}^{+}$, and thus we get
\begin{equation}
\label{ddom2}
{{\cal D}_n}\,\sim \, \frac{1}{(1-z)^{\rho}}
\end{equation}
where $\rho$ is determined by
\begin{equation}
\rho= \sup_{\{k_1 \dots k_n\}} ( k + \sum_{j>\alpha} (j-\alpha)k_j) =
\left\{
\begin{array}{ll}
n/2 \quad & n < 2(\alpha-1) \\
n+1-\alpha \quad & n> 2(\alpha-1)
\end{array}
\right.
\label{rho2}
\end{equation}
which, once we take (\ref{mom-}) into account, yields
\begin{equation}
\label{momfin}
\nu(q)\,=\,
\left\{
\begin{array}{ll}
q/2 \quad & q< 2(\alpha-1) \\
q+1-\alpha \quad & q> 2(\alpha-1)
\end{array}
\right.
\end{equation}
The set of exponents thus has a nontrivial structure,
characterized by a sort of phase transition for $q=2(\alpha-1)$, a
rather universal feature of many systems exhibiting anomalous
transport \cite{V-AD,OTT}. We notice that the parameter ruling the
presence of a phase transition (and the explicit form of the
spectrum) is $\alpha$, that is the exponent describing the
polynomial instability growth of the family of periodic orbits
coming closer and closer to the marginal fixed point, and thus
describing the sticking to the regular part of the phase space: in
the present example $\alpha=\/(\gamma-1)^{-1}$, and thus the
sticking exponent is easily connected to the intermittency index.
We remark that our reference to ``phase transitions" is somehow
arbitrary, since in effect our technique applies in the present
form only to even moments: however results as (\ref{momfin}) are
consistent with all our numerical checks for the whole real $q$
axis.
\subsection{Intermittent diffusion II}

\begin{figure}[h!]
\centerline{\epsfxsize=9cm \epsfbox{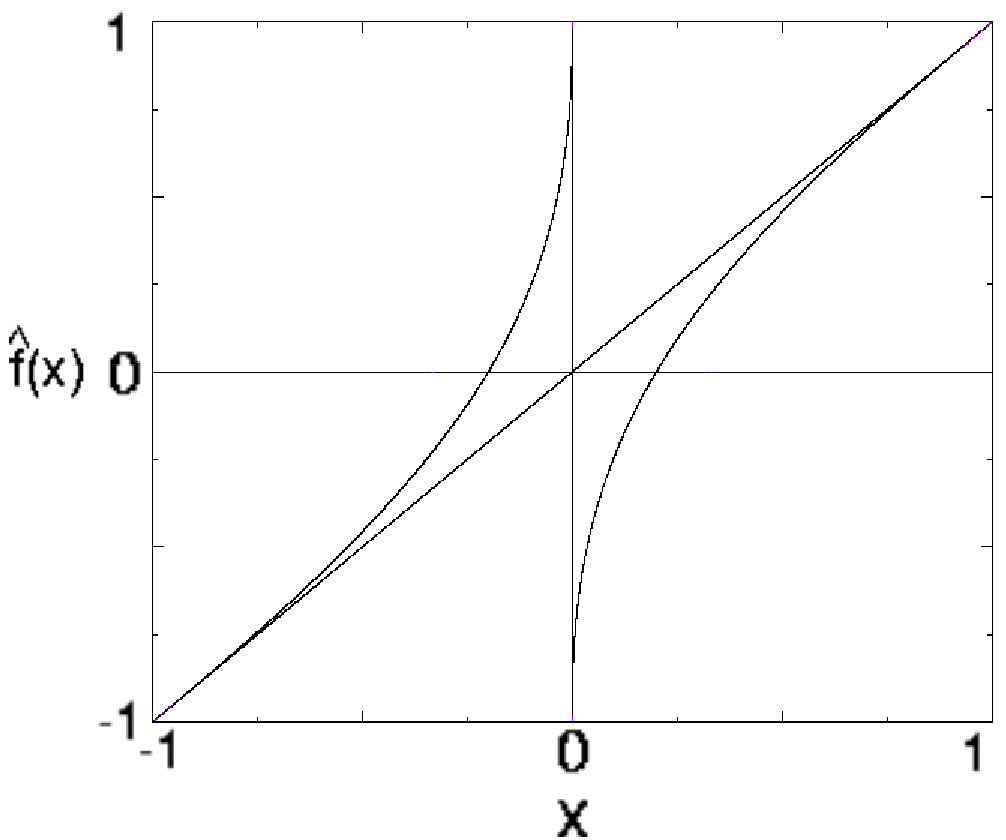}}
\vskip4mm
\caption{\label{pikomap}The intermittent map with constant invariant measure.}
\end{figure}

We now discuss a further example, which will provide additional
evidence of the virtues of our technique. While the setting is
similar to the former case we slightly modify it to conform to
original references. The map we consider belongs to a class of
2-dimensional systems were the diffusive process takes place in a
direction while all the dynamics is limited to the torus on the
other dimension: \ba \left\{
\begin{array}{l}
         x_{t+1}=\hat{f}(x_t)\\
     \\
     y_{t+1}=y_t + x_t
     \end{array}\right.
\ea
where $\hat{f}(x)$ is a  torus map, again dependent on an intermittency parameter $\gamma$,
  implicitly defined on $\mathcal{T}=[-1,1)$ in the following way \cite{Ark}:
\begin{equation}
x\,=\,
\left\{
\begin{array}{ll}
\frac{1}{2\gamma}\left(1+ \hat{f}(x)\right)^{\gamma} \qquad & 0 < x < 1/(2\gamma) \\
\hat{f}(x) + \frac{1}{2\gamma} \left( 1- \hat{f}(x) \right)^{\gamma} \qquad & 1/(2\gamma) < x < 1
\end{array}
\right.
\label{arkmap}
\end{equation}
for negative values of $x$ the map is defined as $\hat{f}(-x)=-\hat{f}(x)$ (see (\ref{symm-P})):
the map is plotted in fig. (\ref{pikomap}).
Diffusion can be treated by the same techniques of the former subsection  if we are able to control the values the function (equivalent to the \emph{jumping numbers} of previous example) 
\beq\label{simjump}
\sigma_p=\sum_{t=1}^{T_p}{x_t}
\eeq
 takes on the periodic orbits.

The most striking difference with respect to the former example is that, {\em for any values of $\gamma$}, the map (\ref{arkmap}) has a {\em constant} invariant measure \cite{Ark}.
\vspace{1cm}

This comes from the summation property
\begin{equation}
\label{arksumm}
\sum_{y=\hat{f}^{-1}(x)}\, \frac{1}{\hat{f}'(y)}\,=\,1
\end{equation}
that easily follows once we observe that each point has exactly two counter images $y_1(x)$ and $y_2(x)$: they come from two different pieces in (\ref{arkmap}) and they have opposite sign: by deriving the two pieces of (\ref{arkmap}) then we get the summation rule.
 This is quite different from the former case where the torus map has non trivial ergodic properties: an absolutely
continuous invariant measure only exists for $\alpha > 1$ (see for instance \cite{ci} and references
therein) while the ergodic behaviour is much more complex when $\alpha <1$. In any case sticking induces
peaking of the measure around marginal fixed points (see \cite{plapp}). In this sense the behaviour of (\ref{arkmap}) is similar to what happens in an area preserving map example \cite{APmap}, where a parabolic fixed point coexist with a Lebesgue invariant measure.

The difference with the previous example  can be euristically understood if we note that now the slope of the map in the chaotic region is not bounded from above: the closer we need to be injected near the marginal points the lesser the probability that this happens. This feature deeply modifies  the instabilities of periodic orbits coming closer and closer to the marginal fixed points and we expect that a linearization as before will not be able to reproduce them.

\begin{figure}[h!]
\centerline{\epsfxsize=10.cm \epsfbox{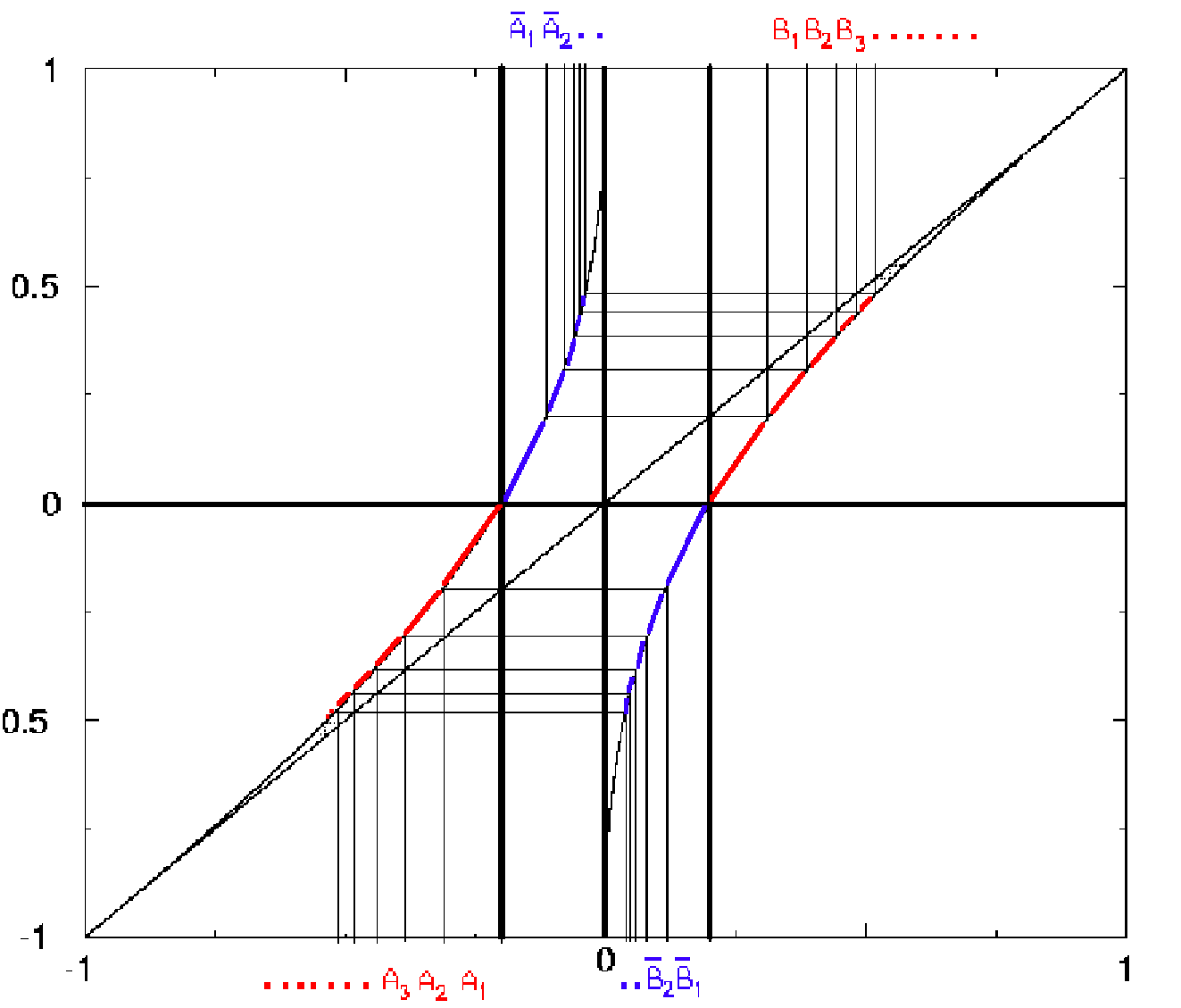}}
\vskip4mm \caption{\label{fpiko} Linearization of the  map
(\ref{arkmap}). Near marginal points, laminar regions, a
Gaspard-Wang type partition is used
while a finer one  is used  in the injection zone
.}
\end{figure}

A linearization is still possible but we must put attention on matching the condition (\ref{arksumm}). To do that let's start by identifying  for each branch a laminar and an injection regions (for example we call $\bar{\rm B}=[0,1/(2\gamma)]$ and ${\rm B}=[1/(2\gamma),1]$ respectively the injection and laminar regions of the right branch (see figure (\ref{fpiko})).

 The Gaspard-Wang  linearization has to be modified in the injection zones ${\rm \bar{A}}$ and ${\rm \bar{B}}$: a linear approximation for all such intervals will give an uniform probability of being reinjected in the laminar zone and this does not respect the peculiar ergodic property of this map. The way out to this problem is to perform a finer  partition of that regions taking the inverse image (inside the injection zones) of each interval partitioning the corresponding laminar zone. For example we will call $\bar{\rm B}_1$ the injection interval corresponding to inverse image of the laminar one ${\rm A}_1$, $\bar{\rm B}_2$ corresponding to ${\rm A}_2$ and so on (see figure (\ref{fpiko})). The linearization is now straightforward for each of this intervals. Moreover the instabilities of all the periodic orbit are well reproduced once we note that the slope of the linearized map restricted to such injection intervals can be safely approximated by (see summation property (\ref{arksumm})):
\beq
\Lambda_{\bar{\rm B}_i}^{-1}=1-\Lambda_{{\rm A}_i}^{-1}
\eeq

We will now show  how  such a linearization is able to reproduce the behaviour of periodic orbit instabilities. We have seen that the  periodic orbits that are relevants for the construction of the zeta function are the ones that accumulate near the marginal fixed point  and for which the instability  has a power law behaviour. 
In this new  map the instability  of such orbits take a linear
correction factor (with respect to the length) from the injection
zone: the intervals in the laminar regions, constructed in the
same spirit of Gaspard-Wang linearization \cite{plapp}, shrink
polinomially with the index $k$: \beq l_{k} \sim
\frac{C}{k^{\alpha+1}}\nonumber \eeq from wich is simple to derive
the periodic orbit instabilities. We rewrite the instabilities of
the relevant orbits for the previous example (I) in comparison
with the ones of this map (II): \ba
\Lambda_{L^kR}^{I}&=&\big[\frac{l_0}{l_1} \frac{l_1}{l_2} \dots \frac{l_{k-1}}{l_k}\big]=\\
&\sim&  k^{\alpha +1}\\
\nonumber\\
\Lambda_{A^kB}^{II}&=&\big[\frac{l_0}{l_1} \frac{l_1}{l_2} \dots \frac{l_{k-1}}{l_k}\big]\big[1-\frac{l_k}{l_{k-1}}\big]^{-1}\big[1-\frac{l_1}{l_0}\big]^{-1}\label{twofact}\\
&\sim&  \Lambda_{L^kR}^{I} \frac{l_{k-1}}{l_{k-1}-l_k}
\sim  k \, \Lambda_{L^kR}^{I}\label{correction}\\
&\sim& k^{\alpha +2}
\ea
where we note that for this second example the modified linearization produce two new factors (see (\ref{twofact})) that correspond to the slope of the two injection intervals visited by the orbit; one of them depends explicitly  on the lenght of the orbit and gives a relevant correction to the instability of the cycle (\ref{correction}).

\begin{figure}[t]
\vskip0.7cm
\centerline{\epsfxsize=8.cm \epsfbox{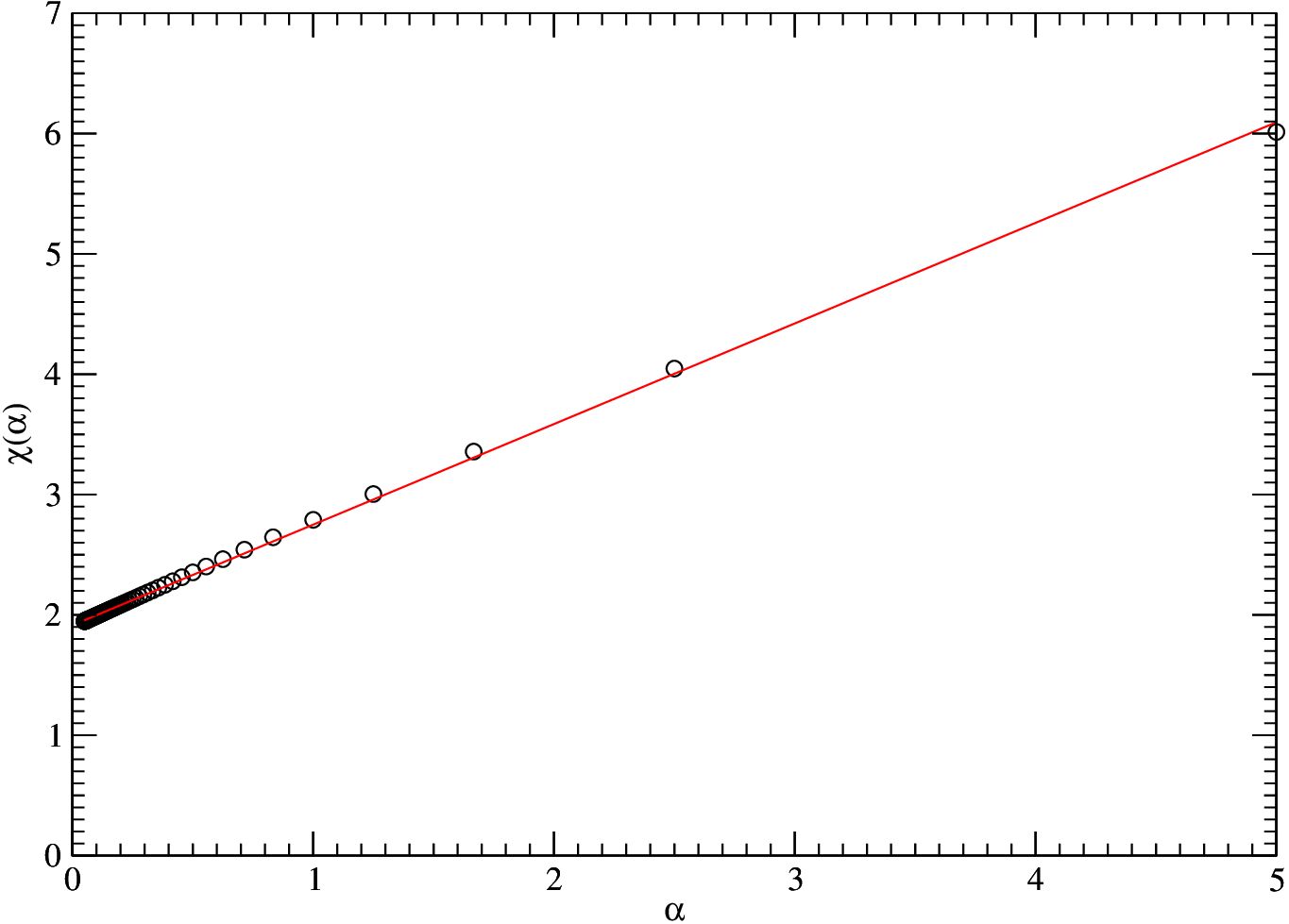}
}
\vskip4mm
\caption{\label{finst}  The exponent $\chi(\alpha)$  of power law decay of instabilities  $\Lambda^{II}_{A^kB} \sim k^{\chi(\alpha)}$ as a function of the parameter $\alpha$ (open circles) and the best fit $\chi(\alpha)=(0.88\alpha+1.9)$ (full
line).
}
\end{figure}

This trend is what we expect from numerical simulations (see figure (\ref{finst})).
Note that this new linearization is able to capture the behaviour of the instabilities of all orbits, not only the one accumulating to the marginal point. As a further example we may consider periodic orbits of the type $A^kB^k$ were the power law exponent  is twice the one before, in agree with performed numerical simulations. 
Thus (\ref{arksumm}) deeply modifies the relationship
between the intermittency exponent and the instabilities of periodic orbits shadowing the
marginal fixed points.

Once we notice that, for the cycles dominating the zeta function expression, the jumping function (\ref{simjump}) can be approximate by the lenght of the periodic orbit (with the appropriate sign), the piecewise linear approximation yields
the dynamical zeta function
\begin{equation}
\zeta_{(0)\beta}^{-1}(z)\,=\,1-\zeta(\alpha+2)z\sum_{k=1}^{\infty}\, \frac{z^k}{k^{\alpha+2}}
\cosh(\beta k)
\label{zark}
\end{equation}
where again $\alpha=1/(\gamma-1)$.

\begin{figure}[h!]
\centerline{\epsfxsize=10.cm \epsfbox{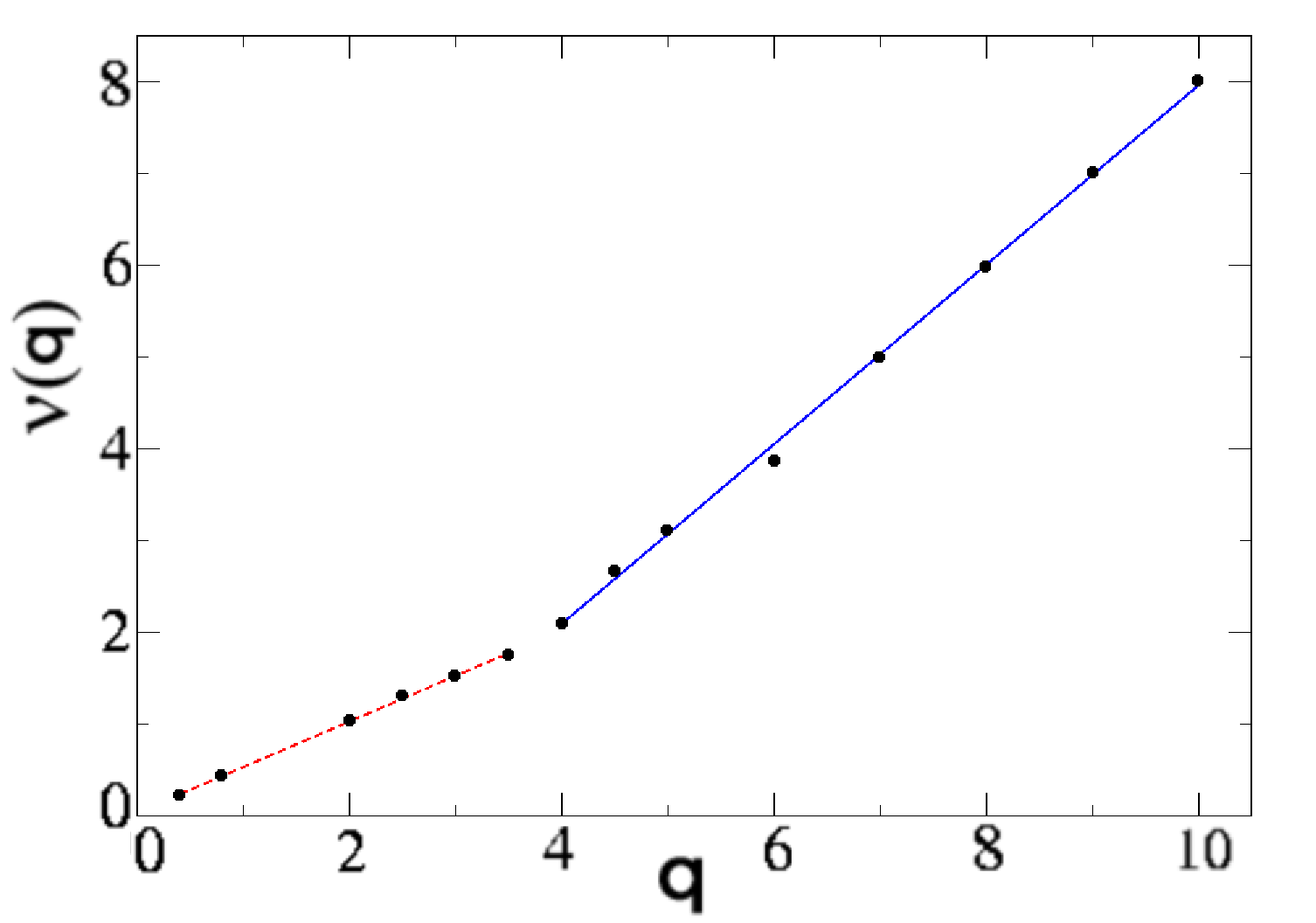}}
\vskip4mm
\caption{\label{spectrum}Spectrum of the transport moments for the map (\ref{arkmap}) with $\gamma=1.5$:
the best fit on numerical data is $\nu(q)=0.50 q +0.04$ for the dotted
line, and $\nu(q)=0.98q-1.82$ for
the full
line.}
\end{figure}

By repeating the steps of the calculation performed in the former
case we get that we always have a phase transition with
\begin{equation}
\label{arkspec}
\nu(q)\,=\,
\left\{
\begin{array}{ll}
q/2 \quad & q< 2\alpha \\
q-\alpha \quad & q> 2\alpha
\end{array}
\right.
\end{equation}
which may also be checked numerically (see figure (\ref{spectrum})).

\subsection{The Lorentz gas with infinite horizon}
\begin{figure}[h]
\centerline{\epsfxsize=10.cm \epsfbox{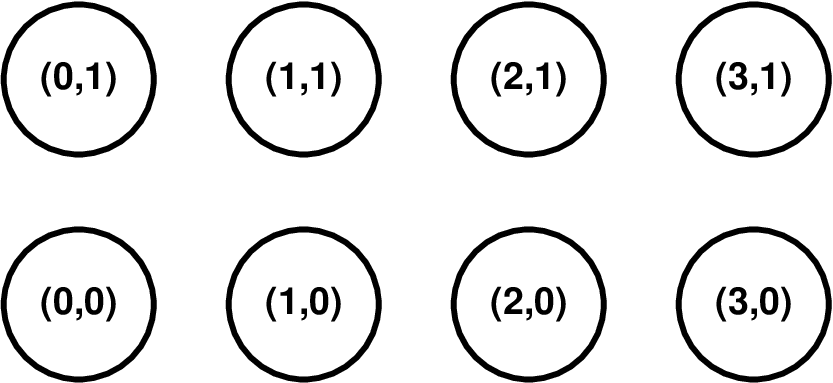}} \vskip4mm
\caption{\label{ilor}Unbounded horizon Lorentz gas, with discs
labelling.}
\end{figure}

The technique does not only apply to one dimensional systems, but is capable of
dealing with more complicated systems, if we are able to characterize the sequence
of orbits that probe more and more accurately the local behaviour near the
ordered region.  A remarkable example in this respect is the Lorentz gas with infinite horizon (see fig. (\ref{ilor})), where a particle collides elastically with a regular array of fixed circular scatterers \cite{LF}, the corresponding torus map being a square billiard table with a circular scatterer in the center (the so called Sinai billiard \cite{LF,A3}). In this example the dynamical feature that lowers the chaoticity of the system is provided by the possibility of infinite free flights along corridors (for instance horizontal and vertical direction in fig. (\ref{ilor})). Periodic orbits for the torus maps (running modes for the full system) that mimic the free flight are for instance orbits that bounce from the $(m,0)$ row to
the $(n,1)$ row travelling a distance of $p$ lattice spacings between each two successive collisions.
In view of our former considerations the estimate we need is how the stability of such orbits scales in the big $p$ limit. Such a calculation is equivalent to computing the fraction of orbits that starting from the $(0,0)$ disc suffer their first collision exactly in the $(p,1)$ disk: such a probability decays like $1/p^3$ (see \cite{per}), and in this sense the Lorentz gas corresponds to the case $\alpha=2$, in (\ref{pl-dz}). This leads to a direct estimate of the spectrum of exponents, according to (\ref{momfin}):
\ba
\label{speclor}
\nu(q) \,=\,\left\{
\begin{array}{ll}
q/2 \quad & q< 2 \\
q-1   \quad & q> 2
\end{array}
\right. \ea It is interesting to show how the phase transition
sits exactly at $q=2$, that is the order corresponding to
diffusion. The behaviour described in (\ref{speclor}), was
obtained recently in \cite{OTT} (actually for a wider class of
extended billiards), by different methods.  We also remark that
our method is also capable of taking track of logarithmic
corrections to the asymptotic estimates, and in the present
example we get that the variance (second moment) asymptotically
grows as $t \cdot \ln t$ (see \cite{A3}). The appearance of a
logarithmic correction may be understood if we notice that, for
$\alpha=2$, we have that
\begin{equation}
\label{secder} \frac{\partial^2}{\partial \beta^2}\,
\zeta^{-1}_{(0)\beta}(\rme^{-s})\, \sim \, \frac {ln
(1-\rme^{-s})}{(1-\rme^{-s})}
\end{equation}
(cfr. (\ref{pl-dz})): then the leading singularity, dominating the
integral (\ref{gen-mom}) is of the form $\ln
(1-\rme^{-s})/(1-\rme^{-s})^2$, which, by using for instance
Tauberian theorems for Laplace transforms, leads to the
logarithmic correction in time. Notice that our results are
asymptotic and they do not depend on the geometric features of the
billiard (radius to cell size ratio), which would of course modify
transient behaviour, as well as prefactors (with dramatic effects
once this ratio tends to $0$ or $1$).
\section{Conclusions}
In this paper we have proposed a zeta function technique to compute the asymptotic behaviour of different transport moments: the essential ingredient in the analysis of weakly chaotic examples, that show deviations from the normal diffusing case, has been shown to be a proper characterization of the sequence of periodic orbits probing closer and closer dynamical features of the marginal structures. In particular the behaviour of the moments' spectrum $\nu(q)$ is governed by the power law relating such orbits' instability to their period. In many cases $\nu(q)$ is characterized by a phase transition between a normal regime and a ballistic one.
\ack
This work was partially supported by INFM PA project {\em
Weak chaos: theory and applications}, and by EU contract QTRANS Network
({\em Quantum transport on an atomic scale}).

\end{document}